# Advanced 3D-Printed Multiphasic Scaffold with Optimal PRP Dosage for Chondrogenesis of BM-MSCs in Osteochondral Tissue Engineering

*Faezeh Ghobadi [1], Maryam Mohammadi [1], Rooja Kalantarzadeh [1], Ehsan Lotfi [2], Shokoufeh Borhan [3], Narendra Pal Singh Chauhan [4], Ghazaleh Salehi [1*], and Sara Simorgh [1,5*].*

[1]Cellular and Molecular Research Center, Iran University of Medical Sciences, Tehran, Iran
[2]Department of Medical Biotechnology, Faculty of Allied Medical Sciences, Iran University of Medical Sciences, Tehran, Iran
[3]Department of Materials, Chemical and Polymer Engineering, Buein Zahra Technical University, Buein Zahra, Qazvin, Iran
[4]Department of Chemistry, Faculty of Science, Bhupal Nobles' University, Udaipur, Rajasthan 313001, India
[5]Department of Tissue Engineering and Regenerative Medicine, Faculty of Advanced Technologies in Medicine, Iran University of Medical Sciences, Tehran, Iran

**\*Corresponding authors:**

**Dr. Sara Simorgh** (ORCID: 0000-0001-6246-9531)

Assistant Professor, Iran University of Medical Sciences, Shahid Hemmat Highway, Tehran, 1449614535, Iran

Tel: (+98 21) 8862 4836; Fax: (+98 21) 8862 2533

E-mail: srsimorgh@gmail.com

**Dr. Ghazaleh Salehi** (ORCID: 0009-0007-7230-0634)

Tel: (+98 912) 4787175;

E-mail: qazalesalehi@gmail.com

**Abstract**

In osteochondral tissue engineering (OCTE), simultaneously regenerating subchondral bone and cartilage tissue presents a significant challenge. Multiphasic scaffolds were created and manufactured using 3D printing to address this issue. Excellent interfacial mechanical properties and biocompatibility enhance the growth and chondrogenic differentiation of bone marrow mesenchymal stem cells (BM-MSCs). The subchondral bone bottom layer is mimicked by incorporating varying concentrations of graphene oxide (GO) (0%, 1%, and 2% w/v) into a bioink composed of alginate (Alg) and gelatin (Gel). Based on evaluations of mechanical and biocompatibility properties, 1% GO is selected for further studies. Subsequently, the GO concentration is kept constant while varying the platelet-rich plasma (PRP) dosage in the multiphasic scaffolds. Different PRP dosages (0%, 1%, 2%, and 3% w/v) are integrated into the Alg-Gel bioink to simulate cartilage tissues. Results indicate that 3D-printed scaffolds containing 1% or 2% PRP exhibit favorable biomechanical properties, with no significant differences observed. However, BM-MSCs exposed to 2% PRP demonstrate enhanced adhesion, growth, and viability. Additionally, real-time PCR and Alcian blue staining confirm increased chondrogenic expression and glycosaminoglycans (GAGs) synthesis. This work highlights the promising potential of 3D-printed multiphasic frameworks in the development of OCTE.

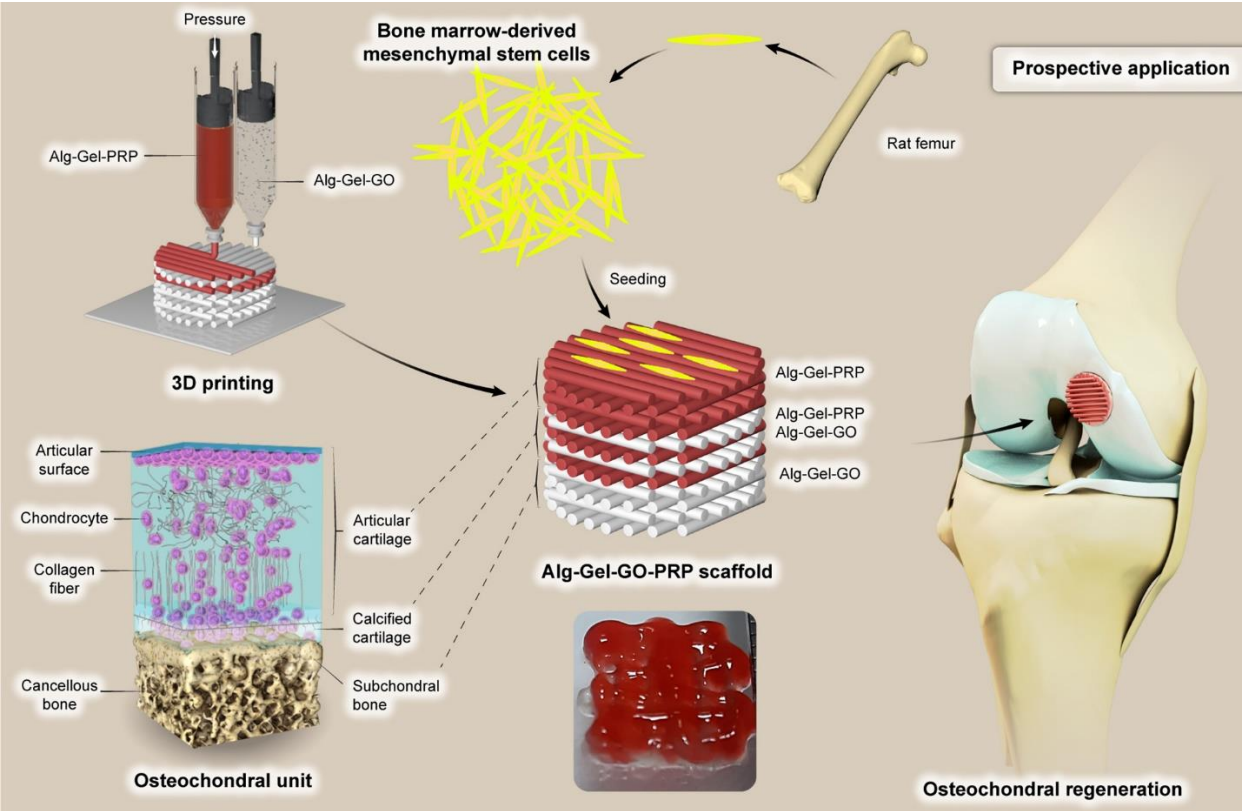

## 1. Introduction

Osteochondral (OC) tissue consists of three key components: articular cartilage, calcified cartilage, and subchondral bone. Each area has a specific function to manage the mechanical forces during joint activities [1]. Repair of OC defects requires simultaneous subchondral bone repair and articular cartilage regeneration [2]. Tissue engineering (TE) strategies for articular cartilage and bone often involve encapsulating cells or growth factors (GFs) in scaffolds or hydrogels, alongside scaffold-free or self-assembly approaches, especially for cartilage. Combining these techniques in multi-layered constructs is essential for effective osteochondral tissue engineering (OCTE) strategies [3,4]. Recent advancements in biological repair methods have introduced scaffolds with intricate structures and unique biomechanical properties, which offer mechanical support for cellular activities and facilitate tissue regeneration [5]. Additive manufacturing, particularly 3D printing, has gained popularity in creating personalized OC scaffolds. The use of multi-material 3D printing, which combines hydrogels, polymers, and cells, has enabled the fabrication of multilayered constructs with precise tissue modeling, allowing for the development of bioinspired scaffolds [6]. This technology enables the creation of scaffolds with customized dimensions, porosities, and mechanical properties, through the use of synthetic and natural biomaterials as bioinks [7].

Natural polymers, such as alginate (Alg), are commonly used in scaffold fabrication because of their biocompatibility, bioactivity, and ease of modification through physical or covalent crosslinking [8]. Alg is commonly employed in 3D printing due to its capacity to transition from a viscous fluid to a stiff hydrogel [9]. However, its limited biocompatibility and bioactivity have led to the incorporation of other bioactive hydrogels, such as gelatin (Gel), to mimic the extracellular matrix (ECM) properties [10]. Alg-Gel blends offer enhanced cell proliferation and better performance compared to pure Alg scaffolds, making them an effective choice for biomedical applications [11]. Despite the advantages of hydrogels, their weak mechanical properties and low processability remain a challenge for 3D printing. To overcome these limitations, reinforced hydrogels incorporating nanomaterials, such as hydroxyapatite, clay, and graphene oxide (GO), have been developed [12]. These nanocomposites improve not only the mechanical and rheological characteristics of hydrogels but also their bioactivity, providing additional capabilities such as optical properties and anisotropic behavior. Among these nanomaterials, GO stands out for its bioactivity and has shown outstanding potential in bone regeneration, particularly when combined with Alg and Gel [13]. The addition of GO into Alg-Gel scaffolds improves their mechanical integrity and biological performance, promoting greater mineralization and mechanical strength [14]. Furthermore, the use of growth factors (GFs), such as those found in platelet-rich plasma (PRP), can accelerate chondrogenesis and stimulate the production of cartilage ECM [15]. PRP is particularly rich in various GFs, such as platelet-derived growth factor (PDGF), vascular endothelial growth factor (VEGF), transforming growth factor beta (TGF-β), basic fibroblast growth factor (BFGF), and epidermal growth factor (EGF) [16]. These GFs play a pivotal role in enhancing the regenerative potential of OCTE strategies, especially by promoting the differentiation of bone marrow-derived mesenchymal stem cells (BM-MSCs) into chondrocytes and supporting cartilage ECM production [17]. Existing studies have not investigated the creation of 3D-printed multiphasic bioactive scaffolds that combine GO and PRP.

This project aimed to create biocompatible, multiphasic, tri-layered scaffolds with distinct mechanical and biological properties. The top layer, designed to mimic articular cartilage, is composed of a blend of Alg and Gel bioink with varying dosages of PRP. The middle layers are made with a layer-by-layer combination of Alg-Gel-PRP and Alg-Gel-GO. The bottom layer, which represents the subchondral bone, incorporates Alg-Gel bioink with an optimum dosage of GO. Up to our knowledge, this is the first study to fabricate a functionally graded OC scaffold using multiphasic Alg-Gel-GO-PRP scaffolds through extrusion-based 3D printing. We subsequently assessed the mechanical and biological performance of a chondroinductive 3D-printed multiphasic scaffold *in vitro* for chondrogenesis in BM-MSCs.

## 2. Experimental Section

### 2.1. Materials

Dulbecco's modified Eagle's medium (DMEM), fetal bovine serum (FBS), penicillin-streptomycin (pen/strep), and 0.25% trypsin-EDTA were obtained from Gibco (USA). All constructs were sustained in a chondrogenic medium comprising high-glucose DMEM enriched with sodium pyruvate, l-proline, l-ascorbic acid-2-phosphate, insulin transferrin-selenium (ITS), and dexamethasone obtained from Sigma Aldrich (USA), and transforming growth factor-β3 (TGF-β3) was purchased from Miltenyi Biotec (Germany). The 3-[4, 5-dimethylthiazole-2-yl]-2, 5-diphenyltetrazolium bromide (MTT) was purchased from Kiazist (Iran). The PRP kit was purchased from Rooyagen (Iran). GO nanopowders (CAS No. 7732-18-5; carbon purity of 99%) was obtained from US Research Nanomaterials (USA). The materials were sourced from Sigma Aldrich (USA): Alg (A2033, medium viscosity, Mw = 120000-190000 g/mol), Gel type B (9000708, 225 bloom), fluorescein diacetate (FDA), propidium iodide (PI), Alcian Blue, glutaraldehyde, and calcium chloride ($CaCl_2$) (CAS No. 10043-52-4/102378). FSP SYBR Green qPCR Mix 2X, TRIZOL reagent, and cDNA synthesis kit were acquired from Fanavari Salamat Partogene (Iran).

## 2.2. Bioink preparation

The OC scaffold was fabricated using a bioink composed of Alg, Gel, GO, and PRP. Two distinct bioinks were formulated: one containing Alg-Gel-GO for the bone phase and another made of Alg-Gel-PRP for the cartilage phase. A uniform solution was obtained by dissolving 10% Alg and 6% Gel in phosphate-buffered saline (PBS). The printing ink for the bone phase was prepared by adding GO nanopowders to the Alg and Gel mixture with varying GO concentrations (0%, 1%, and 2% w/v). Bioinks for the cartilage phases containing PRP were developed by blending Alg-Gel with various PRP concentrations (0%, 1%, 2%, and 3% w/v). **Table 1** provides an overview of the experimental groups, and each group will be referred to using the provided abbreviations.

Table 1: The bioink types and their abbreviations for each phase.

| Tissue Phase | Bioink name | Alg (% w/v) | Gel (% w/v) | GO (% w/v) | PRP (% w/v) |
|---|---|---|---|---|---|
| Subchondral bone | Alg-Gel-$GO_0$ | 10 | 6 | 0 | 0 |
| Subchondral bone | Alg-Gel-$GO_1$ | 10 | 6 | 1 | 0 |
| Subchondral bone | Alg-Gel-$GO_2$ | 10 | 6 | 2 | 0 |
| Calcified cartilage | Alg-Gel-$GO_1$-$PRP_0$ | 10 | 6 | 1 (Optimal dosage) | 0 |
| Calcified cartilage | Alg-Gel-$GO_1$-$PRP_1$ | 10 | 6 | 1 (Optimal dosage) | 1 |
| Calcified cartilage | Alg-Gel-$GO_1$-$PRP_2$ | 10 | 6 | 1 (Optimal dosage) | 2 |
| Calcified cartilage | Alg-Gel-$GO_1$-$PRP_3$ | 10 | 6 | 1 (Optimal dosage) | 3 |
| Articular cartilage | Alg-Gel-$PRP_0$ | 10 | 6 | 0 | 0 |
| Articular cartilage | Alg-Gel-$PRP_1$ | 10 | 6 | 0 | 1 |
| Articular cartilage | Alg-Gel-$PRP_2$ | 10 | 6 | 0 | 2 |
| Articular cartilage | Alg-Gel-$PRP_3$ | 10 | 6 | 0 | 3 |

## 2.3. Multiphasic scaffolds fabrication and characterization

The 3D-printed scaffolds were designed using Solidworks software, with the generated STL files converted to G-code using Simplifier. The lattice structures were then directly fabricated using a commercial 3D bioprinter (BioFabX4, OmidAfarinan Mohandesi Ayandeh, Iran) as part of the printing process. The bioink solutions were loaded into 3D-printing syringes and extruded through a 500-μm nozzle using a pneumatic pressure at 0.6 atm. Each layer of the design consisted of parallel struts oriented at a 90° angle to the layer below. The printing rate was maintained at 3 mm/s, ensuring that struts with a width of approximately 1.6 mm were deposited. The structure designed for printing featured a grid layout with overall dimensions of 9.8 × 9.8 × 1.2 $mm^3$, consisting of four layers of 1.2 mm each designated for cartilage, bone, and an interfacial phase between them, totaling 12 layers for the multiphasic scaffold. Following printing, the scaffolds were immediately soaked in a crosslinking solution consisting of 2% w/v $CaCl_2$ and left at room temperature for 15 minutes to preserve their shape and structural stability. To obtain the best results, printing parameters were changed for each group of multiphasic scaffolds. **Table 2** introduced the names of the various 3D-printed multiphasic scaffold groups. The constant value of GO 1% (w/v) was used in multiphasic scaffold preparation, following optimization in the subchondral layer.

**Table 2: Abbreviations of 3D-printed multiphasic scaffold groups and their bioink compositions in each phase.**

| Multiphasic scaffold groups | Subchondral bone layer | Calcified cartilage layer | Articular cartilage layer |
|---|---|---|---|
| Alg-Gel-GO-PRP$_0$ | Alg-Gel-GO$_1$ | Alg-Gel-GO$_1$-PRP$_0$ | Alg-Gel-PRP$_0$ |
| Alg-Gel-GO-PRP$_1$ | Alg-Gel-GO$_1$ | Alg-Gel-GO$_1$-PRP$_1$ | Alg-Gel-PRP$_1$ |
| Alg-Gel-GO-PRP$_2$ | Alg-Gel-GO$_1$ | Alg-Gel-GO$_1$-PRP$_2$ | Alg-Gel-PRP$_2$ |
| Alg-Gel-GO-PRP$_3$ | Alg-Gel-GO$_1$ | Alg-Gel-GO$_1$-PRP$_3$ | Alg-Gel-PRP$_3$ |

## 2.4. Subchondral bone layer characterization
### 2.4.1. Swelling ratio

To assess the swelling behavior of scaffolds, we initially measured the dry weight of Alg-Gel-GO$_0$, Alg-Gel-GO$_1$, and Alg-Gel-GO$_2$ hydrogels (n = 3) before immersing them in the PBS solution. The scaffolds were removed, dried, and then weighed at different time points to calculate their initial moisture content, otherwise known as their wet weight. The percentage of swelling ratio was calculated using **(Equation. 1)**, where m$_D$ and m$_S$ represent the masses of scaffolds before and after swelling in PBS, respectively.

$$\text{Swelling ratio} = \frac{m_S - m_D}{m_D} \times 100\% \tag{1}$$

### 2.4.2. Mechanical properties

The mechanical properties of Alg-Gel scaffolds containing GO were analyzed using a uniaxial compression test on a SANTAM model STM-20 universal testing machine (Iran). Cylindrical specimens were fabricated, immersed in CaCl$_2$, washed with deionized water, and freeze-dried. A compressive force of 5 kN was applied at a crosshead rate of 1 mm/min to five specimens until they experienced a 60% decrease in height. The compressive strength was determined using **Equation (2)**, where F denotes the greatest force applied and A represents the specimen's cross-sectional area.

$$S = \frac{F}{A} \tag{2}$$

### 2.4.3. Fourier Transform Infrared Spectroscopy (FTIR)

The chemical properties of the Alg-Gel scaffold, prepared with different dosages of GO (0%, 1%, and 2% w/v), were assessed using FTIR (Thermo Nicolet Avatar 360, USA) in the range of 400–4000 cm$^{-1}$. The analysis was conducted under a pressure of 250 Pa, with a resolution of 4 cm$^{-1}$, and it involved averaging 100 scans for enhanced precision.

### 2.4.4. Rheological properties

The rheological properties of Alg-Gel bioink with varying GO concentrations (0%, 1%, and 2% w/v) were assessed using an Anton-Paar Physica MCR300 rheometer with a 25 mm diameter plate-plate geometry. A frequency sweep experiment was carried out by adjusting the frequency from 0.1 to 100 Hz at a constant strain of 1.0%. Additionally, steady-state flow tests were performed in amplitude sweep mode, measuring viscosity across shear rates ranging from 0.01 to 1000 s$^{-1}$. Viscosity recovery tests were conducted in three stages: initially, the viscosity of the bioinks was measured for 130 s while a shear rate of 0.1 s$^{-1}$ was applied for 60 s; subsequently, the shear rate was increased to 100 s$^{-1}$ for 10 s in the second stage; and finally, the shear rate was returned to 0.1 s$^{-1}$ for 60 s in the third stage. This comprehensive analysis aimed to replicate the bioink's behavior before, during, and after the printing process in the first, second, and third stages, respectively. All measurements were performed at a temperature of 25°C.

### 2.4.5. Cell viability assay

The biocompatibility of Alg-Gel bioinks at the concentrations of (0%, 1%, and 2% w/v) of GO was evaluated using the MTT assay. For this purpose, L929 cells (sourced from the Pasteur Institute of Iran) were seeded on sterilized hydrogels in 96-well plates at a density of $5 \times 10^3$ cells per well [18]. The MTT assay was conducted at 1, 3, and 7-day intervals. The culture media were removed, and 10 microliters of MTT reagent, combined with 100 microliters of serum-free media, was added to each well. The plates were stored in the dark at 37°C under conditions of 5% $CO_2$ and 95% relative humidity for 4 hours. Following incubation, the media was discarded, and 100 µL of solubilizing solution was added to each well. The plate was then wrapped in foil and placed on an orbital shaker for 15–20 minutes to ensure complete dissolution of the MTT formazan. A microplate reader (Bio-Rad Laboratories, Hercules, CA) was used to measure absorbance at 570 nm. Viability of cells was assessed by applying **Equation 3**, with a 2D cell culture medium being used as the reference sample for comparison [19].

$$\text{Cell viability\%} = \frac{\text{Mean OD}_{sample}}{\text{Mean OD}_{control}} \times 100 \tag{3}$$

In addition, a field-emission scanning electron microscope (FE-SEM, MIRAIII, TESCAN, Czech Republic) was used to observe cell morphology seeded on the scaffolds [20]. L929 cell lines were seeded on scaffolds in each group at a density of $8 \times 10^3$ cells/well. After 48 hours of incubation, the cells were washed with PBS, fixed with 2.5% (v/v) glutaraldehyde, and then dehydrated using varying concentrations of ethanol (50%, 70%, 90%, 95%, and 100%) for 5 minutes each. Finally, the L929 cell-attached scaffolds were coated with gold-palladium and observed under FE-SEM.

## 2.5. Osteochondral and chondral layer characterization
### 2.5.1. Isolation and characterization of PRP

To prepare PRP, 40 mL of human peripheral blood was collected from consenting healthy volunteers, with 5 mL of acid citrate dextrose solution-A added as an anticoagulant. A 1 mL blood sample underwent a complete blood count. The remaining blood was centrifuged at 1600 rpm for 15 minutes, creating three layers: red blood cells at the bottom, white blood cells in the middle, and plasma on top. The buffy coat and plasma were then centrifuged at 2800 rpm for 7 minutes to concentrate platelets, resulting in 4–6 mL of leukocyte-rich PRP. Platelet counts in the PRP were determined using a hematology analyzer (Sysmex XN 1,000), averaging $540 \times 10^3$ platelets/µL.

### 2.5.2 Lyophilized PRP powder preparation and characterization

The PRP was manually homogenized by inversion, resuspended in 100 mM trehalose, frozen at -80°C, and freeze-dried at -40°C for 24 hours. The final PRP powder was stored at -20°C for further use. The isolated PRP was characterized by enzyme-linked immunosorbent assay (ELISA) and FTIR [21]. The *in vitro* release analysis measured the GFs released from PRP using an ELISA method with specific kits for PDGF-B (R&D Systems, cat. no. DBB00) and TGF-β1 (R&D Systems, cat. no. DY240). The ELISA followed the manufacturer's guidelines, including incubation with assay diluent, control sample addition, washing, and the use of detection antibodies, with optical density measured at 450 nm using a microplate reader. FTIR analysis of PRP powder was performed across a 400–4000 cm$^{-1}$ range, at 250 Pa pressure, with a spectral resolution of 4 cm$^{-1}$ and 100 scans.

### 2.5.3. Printability assessment of bioinks

Shape fidelity was evaluated through qualitative analysis of extruded filaments using macroscopic images and by determining the printability index (Pr) based on the parameter defined by Ouyang *et al.* [22]. To assess printability, 500-μm nozzles were fixed to 10 mL plastic syringes, each filled with 5 mL of a hydrogel blend for testing. The minimum pressure required for consistent extrusion was determined while the nozzle speed and bioink temperature were adjusted to optimize printing precision and duration. The infill density was set at 100% to prevent unwanted porosity in the constructs, and the distance between the needle and the print surface was calibrated to align the leading edge of the flow with the needle. To enhance post-print visualization, two drops of food dye were mixed into the bioinks. The Pr was then calculated by normalizing the pore perimeter by the pore area, as described in **Equation. 4:**

$$\text{Pr} = \frac{L^2}{16A} \quad (4)$$

The perimeter is represented by L, and the area of the enclosed grid hole is denoted by A. Under optimal gelation conditions or ideal printability, the interconnected channels of the constructs would form a perfect square, yielding a Pr value of 1. A higher Pr value indicates a greater degree of bioink gelation, while a lower Pr value signifies a lesser degree of gelation. The Pr value for each set of printing parameters was calculated by analyzing optical images of the printed constructs. The perimeter and area of the interconnected channels were measured using ImageJ software (n=5).

### 2.5.4. Morphological characterization

The surface characteristics and pore arrangement of the manufactured scaffolds were examined and documented using a field emission scanning electron microscope (FE-SEM), with the cross-sections of freeze-dried specimens attached to aluminum foil and then coated with gold. The diameters of the struts and pores were determined using ImageJ software.

### 2.5.5. Physical characterization

The degradation behavior of multiphasic scaffolds with different concentrations of PRP (0%, 1%, 2%, and 3% w/v) was defined in terms of the weight loss in the PBS solution at 37°C for 28 days. After measuring the dry weight ($m_0$), the scaffolds were placed in a PBS solution. The dry weight of the specimens was measured at 7, 14, 21, and 28 days to determine the weight loss of the scaffolds ($m_1$). The degradation rate (D) of each scaffold was calculated according to (**Equation. 5**):

$$D = \frac{m_0 - m_1}{m_0} \times 100\% \tag{5}$$

### 2.5.6. Mechanical and rheological characterizations

The compressive strength of multiphasic scaffolds was measured as described in section 2.4.2. The rheological properties of the multiphasic scaffolds containing varying PRP concentrations (0%, 1%, 2%, and 3% w/v) were evaluated using a rheometer equipped with a plate-plate geometry (25 mm diameter) at 25°C. A frequency sweep was conducted from 0.1 to 100 Hz with a 1% strain amplitude to determine the storage modulus (G') and loss modulus (G″) as functions of frequency. The bioinks' viscosity was measured at a consistent shear rate of 100 $s^{-1}$ to the power of minus one over the entire duration of 200 s.

## 2.6. Biocompatibility characterization of multiphasic scaffold

### 2.6.1. Isolation, expansion, and confirmation of rat BM-MSCs

Using the following technique, BM-MSCs were isolated from the tibiae and femurs of rats [23]. The IUMS Ethical Committee approved the isolation protocols (approval number: IR.IUMS.REC.1401.978). In summary, upon euthanasia, the femora and tibiae were removed. The epiphyses were removed, and the bone marrow was flushed out using DMEM/F12 medium supplemented with 10% FBS, 100 U/mL streptomycin, and 100 µg/mL penicillin. The medium was changed after the cells were cultivated for two days at 37°C with 5% $CO_2$ to eliminate non-adherent cells. Once the cells reached 90% confluence, BM-MSCs were used for the subsequent *in vitro* tests at passage three. Flow cytometry was employed to confirm specific cell surface antigen markers of rat BM-MSCs. Antibodies against positive markers (CD73, CD105, and CD90) and negative markers (CD45 and CD34) were used.

### 2.6.2. MTT and live/dead staining assays

To evaluate the compatibility and cell proliferation of Alg-Gel bioinks with different concentrations of PRP (0%, 1%, 2%, and 3% w/v) and the optimal dosage of GO, an MTT assay was conducted. The assay involved seeding BM-MSCs within sterilized hydrogels in 24-well plates at a density of $10 \times 10^3$ cells per well. After 1, 3, and 7 days, the MTT kit was used as described in section 2.4.4. Cell viability within the 3D-printed scaffolds was assessed through a live/dead assay. BM-MSCs were seeded onto multiphasic scaffolds and cultured in a culture medium. After 72 hours, cell viability was evaluated using FDA and PI [24]. The procedure involved removing the media and rinsing the scaffolds twice with PBS (1X). Subsequently, the staining solution (PI 2 mg/mL, FDA 5 mg/mL, and culture medium without FBS) was applied to the cells and incubated for 10 minutes, and after incubation, the staining solution was removed. An Olympus IX70 fluorescent microscope was used to observe cells, with living cells staining fluorescent green with FDA and dead cells showing fluorescent red with PI in the hydrogel.

### 2.6.3. BM-MSCs morphology and adhesion

FE-SEM was employed to observe the morphology of the cells seeded on the scaffolds. BM-MSCs were seeded onto scaffolds in each group at a density of $3 \times 10^3$ cells/well. After 48 hours of incubation, the stem cells were washed with PBS, fixed with 2.5% (v/v) glutaraldehyde, and dehydrated through a series of ethanol concentrations (50%, 70%, 90%, 95%, and 100%) for 5 minutes each. Finally, the scaffolds with attached stem cells were coated with gold-palladium and examined under FE-SEM.

### 2.6.4. *In vitro* chondrogenesis assay

At passage 3, a BM-MSC suspension with a cell concentration of $3 \times 10^5$ cells/mL was seeded onto the multiphasic scaffold groups in a culture medium (DMEM/F12 supplemented with 10% FBS and 1% penicillin-streptomycin) and incubated for 4 hours at 37°C in a 5% $CO_2$ atmosphere to promote initial adhesion. The control group was prepared by seeding the same number of cells in wells without scaffolds. Finally, the BM-MSCs seeded onto the multiphasic scaffold groups were incubated in a chondrogenic differentiation medium. The chondrogenic differentiation medium, consisting of DMEM with a D-glucose content of 4.5 g/L supplemented with P/S, 120 µM ascorbic acid 2-phosphate, 40 µg/mL L-proline, $10^{-7}$ M dexamethasone, ITS+1 (insulin-transferrin-sodium selenite, linoleic acid-BSA), and 10 ng/mL TGF-β3, was replaced three times per week [25].

### 2.6.5. Gene expression analysis

Following three weeks of incubation in chondrogenic differentiation medium, total RNA was extracted using TRIzol-Chloroform. The mixture was then centrifuged at 12,000 rpm for a period of 15 minutes, with the temperature maintained at 4°C after the addition of chloroform. The samples were subsequently stored at -20°C overnight. The next day, RNA was precipitated and analyzed for quality using a NanoDrop spectrophotometer (Thermo Fisher Scientific, USA). A cDNA synthesis kit was used for complementary DNA (cDNA) synthesis, followed by incubation at 25°C for 10 minutes, 47°C for 60 minutes, and 85°C for 5 minutes in a thermocycler. Reverse transcription-polymerase chain reaction (RT-PCR) was performed using SYBR Premix Ex Taq II master mix (TaKaRa) in a Rotor-Gene Q MDx (USA). Chondrogenic differentiation-related genetic markers such as collagen II, collagen I, and SOX9 were evaluated. The Glyceraldehyde-3-phosphate dehydrogenase (GAPDH) reference gene was used to normalize the target genes, and the run was carried out in accordance with the protocol described by Lotfi *et al.* [26]. The resulting cycle threshold (CT) was converted into $2^{-\Delta\Delta CT}$, and the necessary analyses were performed. The results were reported as relative expression (fold change). The primers used in RT-PCR and their information are listed in **Table 3**.

**Table 3: The primers used in RT-PCR and their information.**

| Gene name | Accession number | Sequence | Product size (bp) | $T_m$ (°C) |
|---|---|---|---|---|
| Collagen type II | NM_001414896.1 | F: 5´-ATCTGTGAAGACCCAGACTGC-3´ | 121 | 60 |
| | | R: 5´-GTTCTCCTTTCTGCCCCTTTGG-3´ | | |
| Sox9 | NM_080403.3 | F: 5´-AGTCGGTGAAGAATGGGCAA-3´ | 158 | 60 |
| | | R: 5´-CTGAGATTGCCCGGAGTGC-3´ | | |
| Collagen type I | NM_053304.1 | F: 5´-GTACATCAGCCCAAACCCCA-3´ | 87 | 60 |
| | | R: 5´-TCGCTTCCATACTCGAACTGG-3´ | | |
| GAPDH | NM_017008.4 | F: 5´-TGTTCTAGAGACAGCCGCAT-3´ | 93 | 60 |
| | | R: 5´-CGATACGGCCAAATCCGTT-3´ | | |

### 2.6.6. Histological analysis

At day 21, Alcian Blue staining was employed to detect the presence of glycosaminoglycans (GAGs) deposits on the multiphasic constructs. At predetermined time intervals, the 3D-printed constructs with BM-MSCs were treated with Alcian Blue solution for staining [27]. Scaffolds were first washed with PBS, fixed in 10% formalin for 30 minutes, and then rinsed. Each scaffold with differentiated cells was incubated with a 1% Alcian Blue solution (in 3% acetic acid, pH 2.5) for 30 minutes. After staining, excess dye was removed, and the scaffolds were rinsed thoroughly. Images were then captured using a Zeiss microscope (Germany).

### 2.7. Statistical analysis

The data are presented as means ± standard deviations from three separate sets of measurements (n = 3). Statistical analysis was conducted using one-way or two-way ANOVA using GraphPad

Prism software. A significance threshold of P < 0.05 was used, with *(P < 0.05), **(P < 0.01), and ***(P < 0.001) indicating varying levels of significance.

## 2.8. Ethical approval statement

The investigation obtained the necessary clearance from the research ethics committee at Iran University of Medical Sciences in Tehran, Iran, with the Ethical Code being IR.IUMS.REC.1401.978. All experiments adhered to relevant guidelines and regulations, which were endorsed by the Research Ethics Committee of Iran University of Medical Sciences, located in Tehran, Iran.

## 3. Results and discussion
### 3.1. Subchondral bone layer characterization
### 3.1.1. Physicochemical analysis of the subchondral layer

**Swelling ratio**

Bioinks of the subchondral layer with different GO concentrations (0%, 1%, and 2% w/v) were prepared as described previously **(Figure. 1a)**. The ability of a material to absorb water is a key factor to take into account, because cells take in nutrients from the surrounding environment to sustain their mobility, growth, and proliferation. This is also related to the functioning of the scaffold structure and the attainment of a precise spatial distribution of cells [28,29]. According to the results **(Figure. 1b)**, the Alg-Gel-GO0 scaffold exhibits a notably greater swelling ratio compared to the Alg-Gel-GO1 and Alg-Gel-GO2 scaffolds. The GO-free scaffold showed 1700% swelling in PBS, making it difficult to handle due to its weak mechanical properties. In GO-containing scaffolds, the swelling ratio after 4 hours of immersion was lower than that of GO-free scaffolds, with values of 1200% and 1530% for Alg-Gel-GO$_1$ and Alg-Gel-GO$_2$, respectively. As GO is a hydrophobic, scaffolds containing it exhibit a lower water absorption capacity compared to those without it. Similar findings regarding the swelling ratio with GO addition were reported by Hu *et al*. [30]. Water uptake in Alg-Gel-GO$_2$ increased as the GO content increased after 4 hours. The mechanical and structural stability of Alg-based printed scaffolds is critical. Divalent ions, or ionic crosslinkers, can separate from ionically cross-linked Alg hydrogels, decrosslinking the scaffolds and resulting in a loss of structural stability [31]. According to the findings, incorporating

GO into Alg helps maintain the structural integrity of Alg-based scaffolds. Good interfacial adhesion between GO and Alg chains, along with hydrogen bonding, is likely to lead to higher scaffold stability. In an investigation by Choe *et al*. [31], a lack of $Ca^{2+}$ ions caused swelling in Alg scaffolds with an Alg content of 3%; at the same time, GO into Alg decreased the swelling ratio and improved the structural stability of Alg-based scaffolds within 5 days of incubation. A higher GO content in the Alg-Gel-$GO_2$ scaffold caused more water molecules and hydrophilic protein uptake, leading to a higher swelling ratio. Consequently, the interaction between GO and Alg polymer chains enhances the mechanical integrity of the printed scaffolds at a GO content of approximately 2.0 mg/mL, thereby improving structural stability.

**Mechanical properties**

The mechanical strength, a crucial aspect in the fabrication of scaffolds for bone TE, indicates the load-bearing capacity of the materials [32]. According to the results **(Figure. 2c)**, the addition of GO obviously improved the compressive strength (0.72 MPa for 1 wt% and 0.63 MPa for 2 wt% GO) compared to ink without GO (0.55 MPa). The 3D-printed scaffolds with Alg-Gel-$GO_0$ exhibited brittleness, making them difficult to handle, with edge sections detaching due to weak mechanical integrity. According to the results, incorporating GO into Alg can improve scaffolds' structural stability. The increase in mechanical behavior as the GO content increases is attributed to the further formation of bridges through GO nanofillers, which reinforces the 3D structure of the scaffolds [12]. On the other hand, interactions such as hydrogen bonding between the polymer matrix (Alg chains) and the GO nanosheets, due to their good compatibility, increased in compressive strength [29]. The improved strength of the GO inks in comparison to the inks without GO highlighted the reinforcement effect of GO incorporation. As shown in **Figure. 2c**, the scaffold with 1 wt% GO represents a higher compressive strength than that of 2 wt% GO, revealing a non-linear relationship between enhancing the mechanical properties of the scaffold and an increase in the GO concentration. When the incorporation of GO is excessively high, the structure is susceptible to collapse, likely due to the swelling and high-water absorption capacity of the GO, which reduces the structure's stiffness. Compressive stiffness is closely linked to water content; as water content rises, the structure becomes more brittle and permeable [29]. To enhance the mechanical properties of the scaffold an optimal amount of GO should be added.

**FTIR analysis**

FTIR spectra of Alg-Gel-GO$_0$ and Alg-Gel-GO$_1$ hydrogels are shown in **Figure. 1d**. Since the FTIR spectra of Alg-Gel-GO$_1$ and Alg-Gel-GO$_2$ do not show differences between the peaks, the spectra of Alg-Gel-GO$_2$ are not shown in the FTIR figure. The band at 1027 cm$^{-1}$ represents a characteristic peak for Alg, confirming its glucuronic acid unit [33]. The peak associated with carboxyl groups present in the Alg is observed at 1409 cm$^{-1}$ [34]. A prominent absorption peak occurs at 1614 cm-1, which is associated with the asymmetric stretching vibrations of the -COO$^-$ group [35]. For Gel, the characteristic peaks at 1544 and 1237 cm$^{-1}$ are attributed to Amide II and Amide III, respectively [34]. These characteristic peaks were also observed in the spectra of Alg-Gel-GO$_1$. The FTIR analysis revealed a characteristic peak for GO at 1777 cm$^{-1}$ in the spectra of the Alg-Gel-GO$_1$. The region spanning 1700 to 2200 cm$^{-1}$ is associated with the C=O stretching motions of the − COOH group, which can be ascribed to C=O linkages found within carbonyl components [34,35]. The peak around 1095 cm$^{-1}$ represents the C–OH bending mode of GO [36]. Furthermore, the interaction between GO and Alg-Gel hydrogel led to the broadening of the peak occurring at 2900-3600 cm$^{-1}$ [13].

**Rheological properties**

A proper viscosity is crucial for successful printing, and assessing the rheological behavior of biomaterials can help identify the optimal printing characteristics [37]. The viscosity-shear rate curves for bioink formulations with varying GO content are shown in **Figure. 2e**. GO-containing bioinks exhibited higher viscosities than the GO-free bioink (Alg-Gel-GO$_0$). Incorporating GO into the Alg-Gel bioink increased its viscosity and enhanced its shear-thinning behavior, making it well-suited for 3D printing. A similar finding was reported by Zhang *et al*. [29], who observed that viscosity increased with GO concentration in the Alg-Gel bioink. A common characteristic observed across all tested bioinks was their shear-thinning behavior, with a consistent decrease in viscosity as the shear rate was raised. This shear-thinning property is vital for bioinks to be effectively extruded from the nozzle at low pressure while preserving the printed structure's shape and dimensions post-printing [31]. The bioink viscosities decreased with higher GO concentrations, with measurements at a shear rate of 1 s$^{-1}$ showing viscosities of 92, 379, and 376 for Alg-Gel-GO$_0$, Alg-Gel-GO$_1$, and Alg-Gel-GO$_2$, respectively. Frequency sweep tests revealed that all bioinks exhibited a larger storage modulus (G') than loss modulus (G"), indicating solid-like behavior and suitability for printing **(Figure. 2f)**. The addition of GO to the Alg-Gel bioink improved the storage modulus, suggesting enhanced structural integrity. GO incorporation also increased both the storage and loss moduli, indicating improved viscoelastic properties, printability, and shape fidelity, as the interactions between Alg and GO resulted in a more viscous and stable hydrogel. Increasing the GO concentration to 2% resulted in a reduction of both moduli, with the highest modulus recorded at a concentration of 1%. The thixotropic behavior of the hydrogels is a key factor in printability and resolution of 3D-printed structures [12]. **Figure. 1g** shows the viscosity recovery behavior of Alg-Gel-GO bioinks. For Alg-Gel-GO$_0$, the initial viscosity was 595 Pa.s. Upon increasing the shear rate to 100 s$^{-1}$ (Step I), the viscosity sharply decreased to 4.54 Pa.s. After removing the shear rate (Step II), the viscosity recovered to 236 Pa.s within 60 seconds, reaching 39.7% of the initial value. In contrast, Alg-Gel-GO$_1$ and Alg-Gel-GO$_2$ bioinks recovered 98.54% and 82.37% of their initial viscosity, respectively. Both of these bioinks exhibited a smaller viscosity decrease in Step II compared to Alg-Gel-GO$_0$, though this did not significantly affect their printability due to similar viscosities. The higher viscosity of Alg-Gel-GO$_0$ led to a lower recovery value. The incorporation of GO significantly improved viscosity recovery, indicating enhanced viscosity and stability. These rheological properties ensure the layer-by-layer integrity of 3D-printed scaffolds, preventing collapse or destruction during printing.

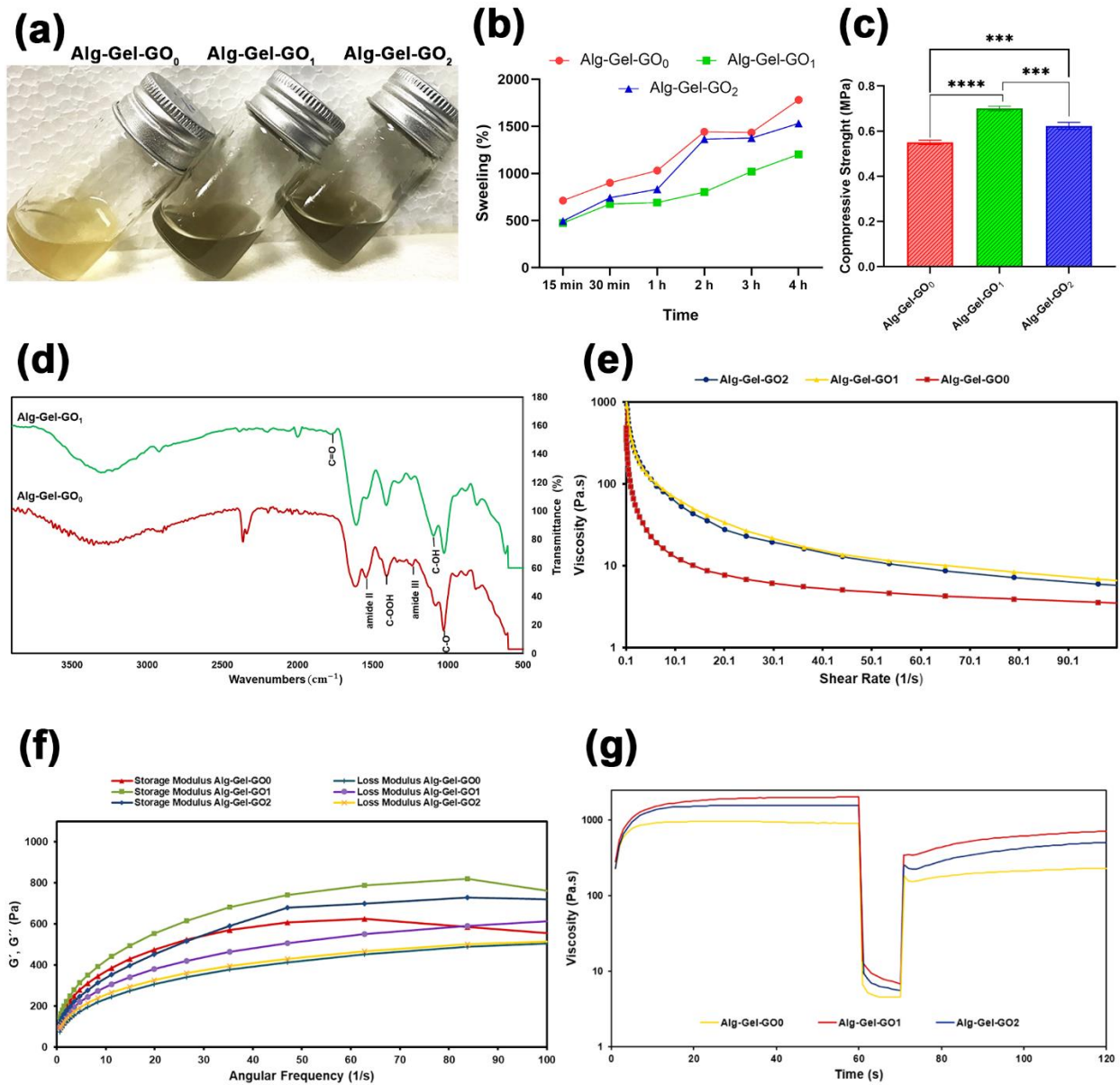

**Figure. 1.** (**a**) The image of the subchondral layer bioinks with different GO concentrations. (**b**) Swelling ratio in PBS as a function of immersion time of scaffolds with various GO concentrations. (**c**) Compressive strength of bioinks with different GO content; (ns, no significant difference; $^*p \leq 0.05$, *** $p \leq 0.001$, **** $p \leq 0.0001$). (**d**) FTIR spectra of Alg-Gel-$GO_0$ and Alg-Gel-$GO_1$ bioinks. Rheological properties of GO-containing hydrogels: (**e**) Viscosities versus different shear rates, (**f**) Dependence of storage modulus (G') and loss modulus (G") on different angular frequencies, and (**g**) Recovery behavior.

### 3.1.2 Biocompatibility analysis of subchondral layer

The impact of GO concentrations on cell viability and morphology was evaluated using MTT and FE-SEM analyses, respectively. In **Figure. 2a,** the MTT results indicate L929 fibroblast cell viability in hydrogels on days 1, 3, and 7. The results indicate that all Gel-Alg-GO groups exhibited non-cytotoxic effects over the 7 days. Cell viability was observed across all groups, with increased concentrations of GO resulting in a decrease in cell death, showing significant variations between the Alg-Gel-$GO_1$ and Alg-Gel-$GO_2$ groups compared to the Alg-Gel-$GO_0$ group on days 1 and 7. This finding aligns with other studies that reported no adverse effects from scaffolds containing GO. Similarly, researchers observed no cytotoxicity at low GO concentrations (1–3 µg/mL) [38,39]. However, higher concentrations (≥100 µg/mL) have been shown to induce cytotoxicity, as demonstrated by Lim *et al.* [40]. Furthermore, the data demonstrated that the cell proliferation rate was significantly higher over time in all Alg-Gel-GO groups, suggesting that these scaffolds effectively support cell proliferation. However, after 7 days, a slight decrease in cell viability was observed in the Gel-Alg-$GO_2$ group compared to the Gel-Alg-$GO_1$ group. Our results were similar to the findings of Zhang *et al*. [29], who developed a novel MSC-laden Alg-Gel-GO composite bioink and noted improved cell viability with higher GO concentrations. Higher GO concentrations (1% and 2% GO w/v) supported earlier cell spreading by day 7 compared with 0.5% and 0% GO. However, cell proliferation decreased at the 2% GO concentration over time. Moreover, the attachment of L929 fibroblast cells to the scaffolds was analyzed using FE-SEM (**Figure. 2b**). The cells displayed good spreading and elongation on the subchondral layer scaffolds, especially with a 1% increase in GO concentration. The morphological analysis results align with the MTT assay findings and previously reported literature [41,42]. This enhanced cell attachment is likely due to the presence of negatively charged carboxylic, hydroxyl, and epoxy groups in the Alg-Gel-GO nanocomposite scaffolds [43]. Based on the MTT and cell attachment test results, the Alg-Gel-$GO_1$ group appears to be a promising candidate for enhancing cell proliferation and adhesion.

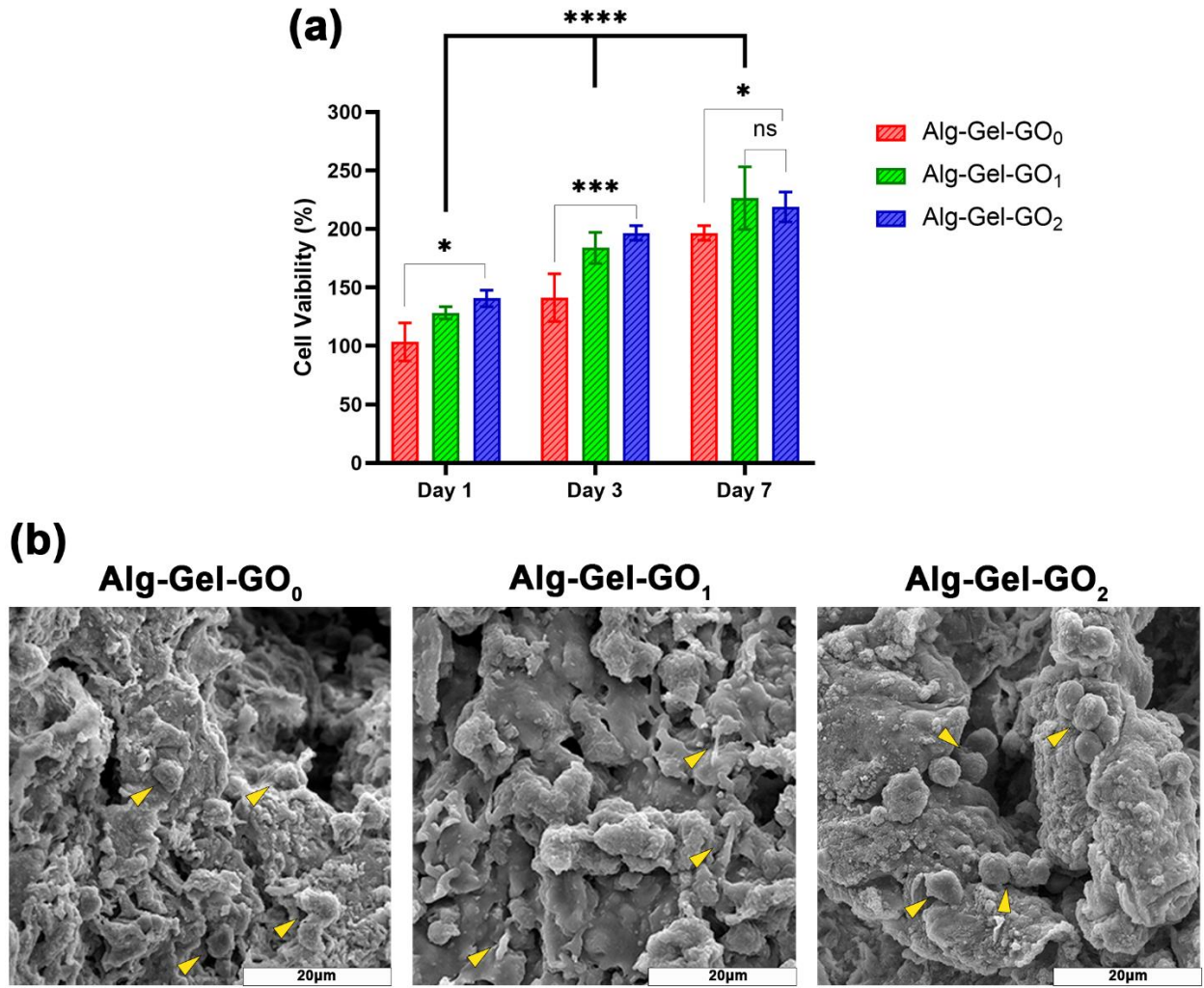

**Figure. 2.** Evaluation of seeded cell viability on scaffolds **(a)** Cell viability across different concentrations of GO was compared using the MTT assay; Data expressed as mean ± SD (n = 3); ns: no significant difference, *p < 0.05, **p < 0.01, *** p < 0.001, **** p ≤ 0.0001. **(b)** FE-SEM images of the adhesion of L929 cells on the surface of Alg-Gel-$GO_0$, Alg-Gel-$GO_1$, and Alg-Gel-$GO_2$.

## 3.2 Osteochondral and chondral layer characterization
### 3.2.1 Lyophilized PRP powder characterizations

**Figure. 3a** shows the isolated PRP and **Figure. 3b** shows the lyophilized PRP powder. The kinetics of TGF-β1 and PDGF release from the lyophilized PRP powder were evaluated using an ELISA. The findings demonstrated that TGF-β and PDGF concentrations increased with PRP concentration and that GFs were released from PRP roughly linearly (**Figure. 3c-d**). These factors promote the growth and proliferation of bone and cartilage cells in the OC area, facilitating tissue regeneration [44]. The preparation of lyophilized PRP powder was confirmed by FTIR, as displayed in **Figure. 3e**. The peak at 1662 cm$^{-1}$ is related to the amide I band in PRP powder [45,46]. Furthermore, the detected peak at 1546 cm$^{-1}$ corresponded to the amide II functional group with vibration and bending of the v (N–H) bond [47]. The bands around 1250–1500 cm$^{-1}$ are possibly assigned to the vibrations of amine groups [48].

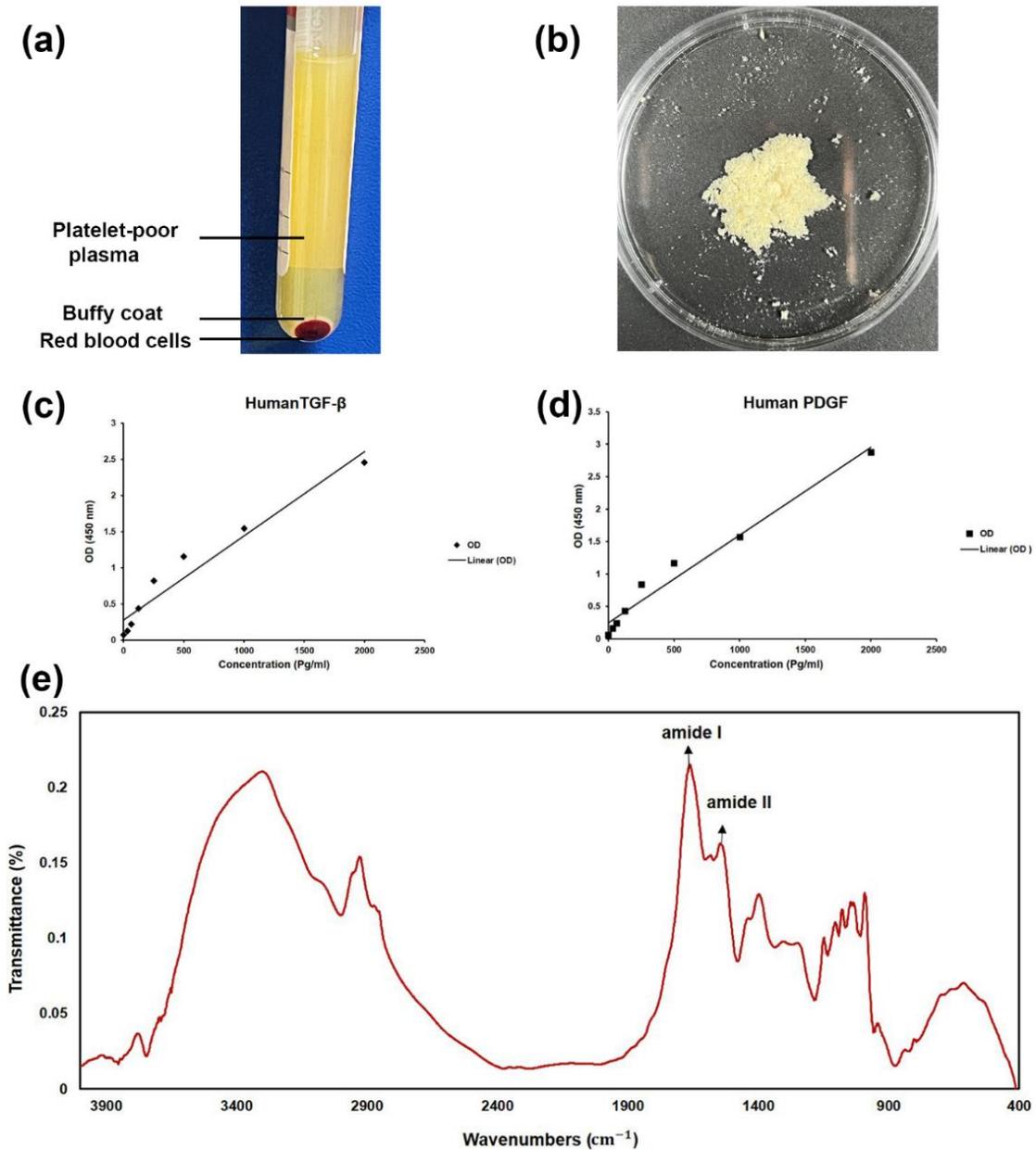

**Figure. 3. (a)** Isolated PRP from human peripheral blood. **(b)** Lyophilized PRP powder. Release of GFs, including **(c)** PDGF and **(d)** TGF-β1 from PRP. **(e)** FTIR spectra of PRP powder.

### 3.2.2 Morphological characterization

Pre-designed CAD models of the 3D-printed scaffolds are shown in **Figure. 4a-c**. Using the 3D printer, the multiphasic scaffolds with a hierarchical structure were successfully fabricated (**Figure. 4d**). The FE-SEM analysis of the printed structure showed an open, porous design that facilitates efficient nutrient exchange (**Figure. 4e**). The structure has an open-space macro grid layout with interconnecting pores in each grid or 3D-printed filament to facilitate cellular interactions. The highly interconnected porosity of a scaffold is crucial for supporting optimal cell seeding, adhesion, and migration. A high level of porosity in the hydrogel is essential for facilitating the transport of nutrients and oxygen within the 3D construct, especially when a functional vascular system is lacking [49]. The cross-sectional morphology of lyophilized multiphasic scaffold groups, including Alg-Gel-GO-PPR$_0$, Alg-Gel-GO-PPR$_1$, and Alg-Gel-GO-PPR$_2$ was observed by FE-SEM (**Figure. 4f-h**), and the pore size was determined by analyzing 10 randomly selected FE-SEM images for each sample using ImageJ software. An interconnected porous structure with a characteristic pore size ranging from 10 to 110 µm was observed in all samples. Moreover, the pore size is crucial for the scaffold's properties, as the pores support cellular proliferation and migration. Previous studies have indicated that a minimum pore size of around 100 µm is necessary for OC tissue regeneration, as it creates hypoxic conditions that promote osteochondrogenesis rather than osteogenesis [50]. **Figure. 4i** shows the pore size values of the multiphasic scaffolds. The pore size of the scaffolds did not change significantly with the addition of increasing concentrations of PRP to Alg-Gel-GO bioink.

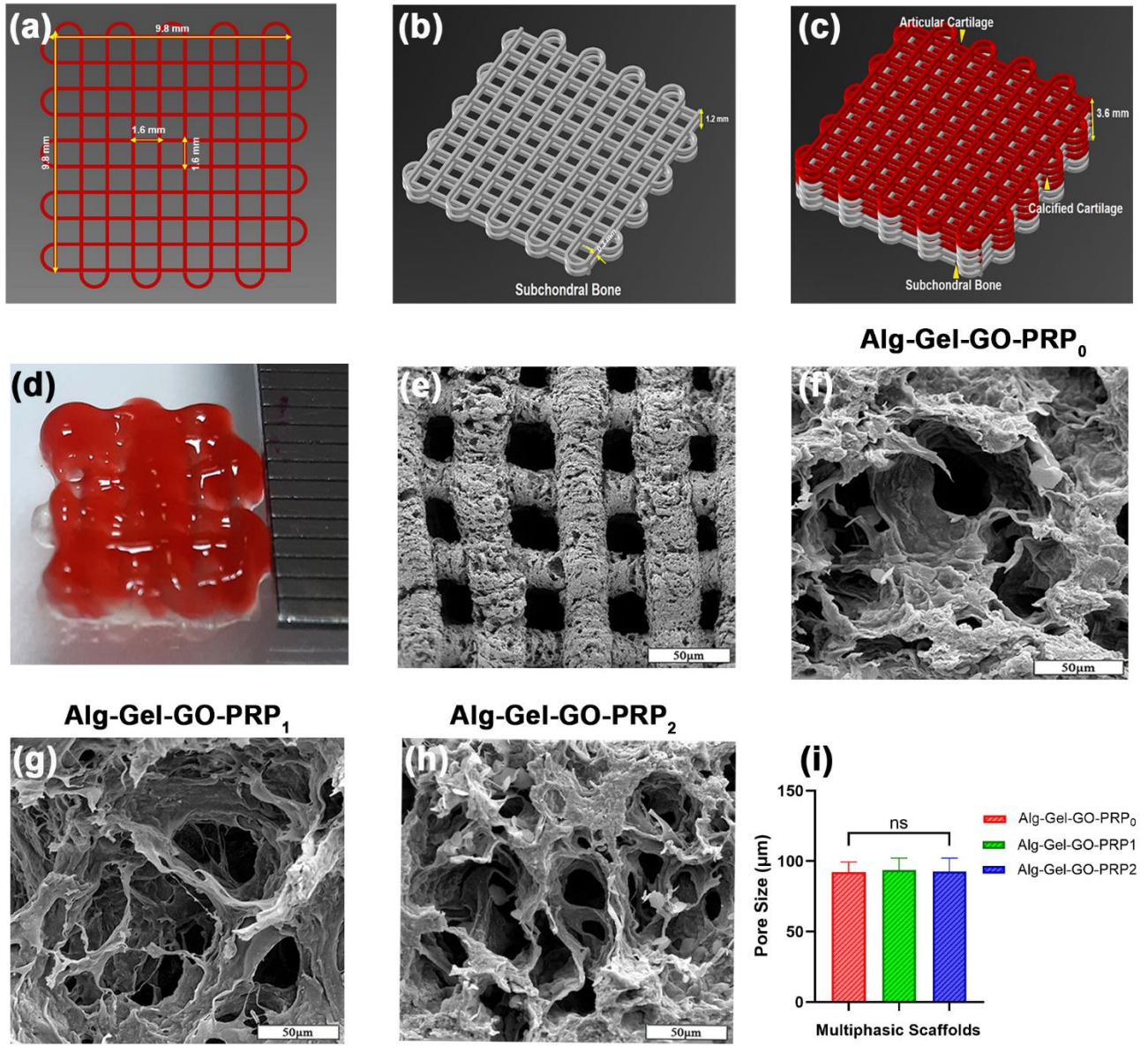

**Figure. 4.** Design and printing of multiphasic scaffolds. (**a**) The single-layer structure is designed with a grid arrangement. (**b**) The subchondral bone layer is constructed. (**c**) The printed structures display 12 layers, which are made up of subchondral bone, calcified cartilage, and articular cartilage. (**d**) A view of the 3D-printed multiphasic scaffold is presented. (**e**) FE-SEM images of the printed structure demonstrated the porous grid structure. FE-SEM images of (**f**) Alg-Gel-GO-PRP$_0$, (**g**) Alg-Gel-GO-PRP$_1$, and (**h**) Alg-Gel-GO-PRP$_2$. (**i**) Pore size values of the multiphasic scaffolds; ns: no significant difference.

### 3.2.3. Printability, degradation rate, rheological, and mechanical properties

The printability of the bioink was evaluated through techniques like rheological analysis and assessment of the integrity of the printed multilayer construct. **Figure. 5a** illustrates the four-layered 3D-printed structure, including Alg-Gel-GO$_1$, Alg-Gel-PRP$_0$, Alg-Gel-PRP$_1$, Alg-Gel-PRP$_2$, and Alg-Gel-PRP$_3$. However, we encountered some challenges with the Alg-Gel-Go-PRP$_3$ sample, as it wasn't suitable for printing, and we couldn't create a scaffold with the PRP 3% (w/v) concentration. Because of this, we decided it would be best to exclude this sample from our further characterization efforts. Printing the bioink under optimal gelation conditions produced smooth, uniform filaments that were continuously extruded, forming a well-defined grid structure with distinct, clearly separated layers [22]. To semi-quantitatively assess the printability of the bioink in this study, Pr was calculated for each parameter combination using **(Equation. 4)**. According to Ouyang's findings [22], a Pr value for Alg-Gel bioink in the range of 0.9–1.1 indicates that the 3D-printed hydrogel construct will exhibit excellent filament morphology and mechanical stability. According to **(Figure. 5a)**, the Pr value of Alg-Gel-GO$_1$ was 0.85±0.06, which had the highest viscosity. Our findings indicate that adding GO to the bioinks significantly enhanced the shape fidelity and resolution of the 3D-printed scaffolds, primarily due to the quicker viscosity recovery after the bioink was extruded [12]. Alg-Gel-PRP$_0$ and Alg-Gel-PRP$_1$ were printable, with Pr values of 0.83±0.01 and 0.8±0.04, respectively, and produced smooth, cohesive grid structures. In contrast, the grids of Alg-Gel-PRP$_2$ exhibited a crooked and uneven shape. It was clear that although Alg-Gel-PRP$_2$ with a Pr value of 0.76±0.04 was printable at 25°C, it lacked sufficient viscosity to preserve the grid structure. In this regard, the concentration of PRP (0%, 1%, and 2% w/v) and proportion were selected due to the printability of the inks in our experiment **(Figure. 5b)**.

The scaffold degradation could cause greater cell exchange between the in situ autologous tissue and the printed structure, allowing more accurate delivery of the seeded cells or bioactive molecules to the damaged site. Thus, the degradation behavior of biomaterials is an essential factor for TE applications [51,52]. The results showed that all scaffolds were biodegradable and that the weight loss of the groups increased over time and decreased with the incorporation of PRP in Alg-Gel scaffolds. The formation of a fibrin network within the constructs, driven by the crosslinking of fibrinogen in PRP, could explain this phenomenon [53]. As a result of biodegradation analysis, the degradation rate of free-PRP scaffolds over 28 days was relatively higher than that of PRP ones, and the mass loss percentage was 78%. With the addition of PRP, the mass loss percentages decreased to 60% and 64% for groups of Alg-Gel-GO-PRP$_1$ and Alg-Gel-GO-PRP$_2$, respectively. In an investigation by Zhao *et al.* [51], an increase in the PRP concentration led to an increase in the remaining mass percentages of Alg-Gel composite hydrogel bioinks. They also exhibited a slower degradation rate in groups with higher PRP concentrations compared to those with lower PRP concentrations. **Figure 5c** illustrates that the Alg-Gel-GO-PRP$_2$ scaffold degrades slightly more quickly than Alg-Gel-GO-PRP$_1$, largely due to lower crosslinking and the formation of a structure that allows for greater water uptake, resulting in increased degradability. However, Alg-Gel-GO-PRP$_1$ and Alg-Gel-GO-PRP$_2$ demonstrated relatively similar degradation profiles, and no significant difference in weight loss was observed within soaking days. The *in vitro* degradation rate of all scaffolds could last more than four weeks, supporting the regeneration of cartilage [53].

Mechanical tests revealed that the compressive strength of multiphasic scaffolds increased with the incorporation of PRP **(Figure. 5d)**. The compressive strength of the 3D-printed scaffolds with PRP (1.15 MPa for Alg-Gel-GO-PRP$_1$ and 1.01 MPa for Alg-Gel-GO-PRP$_2$) was higher than that of the scaffolds without PRP (0.88 MPa), which was in line with the results of the degradation study [53]. The mechanical strength of Alg-Gel-GO-PRP$_2$ is slightly lower than that of Alg-Gel-GO-PRP$_1$, but the difference is insignificant. The ultimate compressive strength of cancellous bone ranges from 0.7 to 30 MPa, with cellular sawbones exhibiting values between 1.4 and 5.4 MPa, and trabecular bone structures ranging from 0.7 to 30 MPa [54]. The compressive strength of the multiphasic scaffolds fulfilled the requirements of a porous scaffold for OC regeneration.

The frequency sweep test showed that the storage modulus (G') value was larger than the loss modulus (G″) value in the whole angular frequency range in both hydrogels **(Figure. 5e)**, demonstrating that the elastic property dominated the viscous property while applying a load. This behavior indicates a typical gel structure formation, characterized by a solid-like behavior in hydrogels [53,55]. The storage modulus was around 630 Pa in Alg-Gel-PRP1 hydrogel, which was relatively higher than that of PRP$_2$ hydrogel (570 Pa) at the frequency of 10 s$^{-1}$, but a significant difference was not observed between Alg-Gel-PRP$_1$ and Alg-Gel-PRP$_2$. This reason may be due to an ionic imbalance inside the hydrogel structure followed by PRP incorporation, and therefore a decrease in stability was observed in hydrogel. Furthermore, the G′ amount is associated with the crosslink density, and the higher the G′ of the hydrogel, the greater the crosslink density [45,56].

Accordingly, the higher G′ of Alg-Gel-PRP$_1$ compared to Alg-Gel-PRP$_2$ may be due to the higher cross-linking density of Alg-Gel-PRP$_1$ than that of Alg-Gel-PRP$_2$. The thixotropic or shear-thinning nature of the hydrogels was shown in **Figure. 5f**, as the viscosity decreased with time under a constant shear rate of 100 s$^{-1}$. This property is particularly beneficial for facilitating the extrusion of highly viscous hydrogels through the printing nozzle.

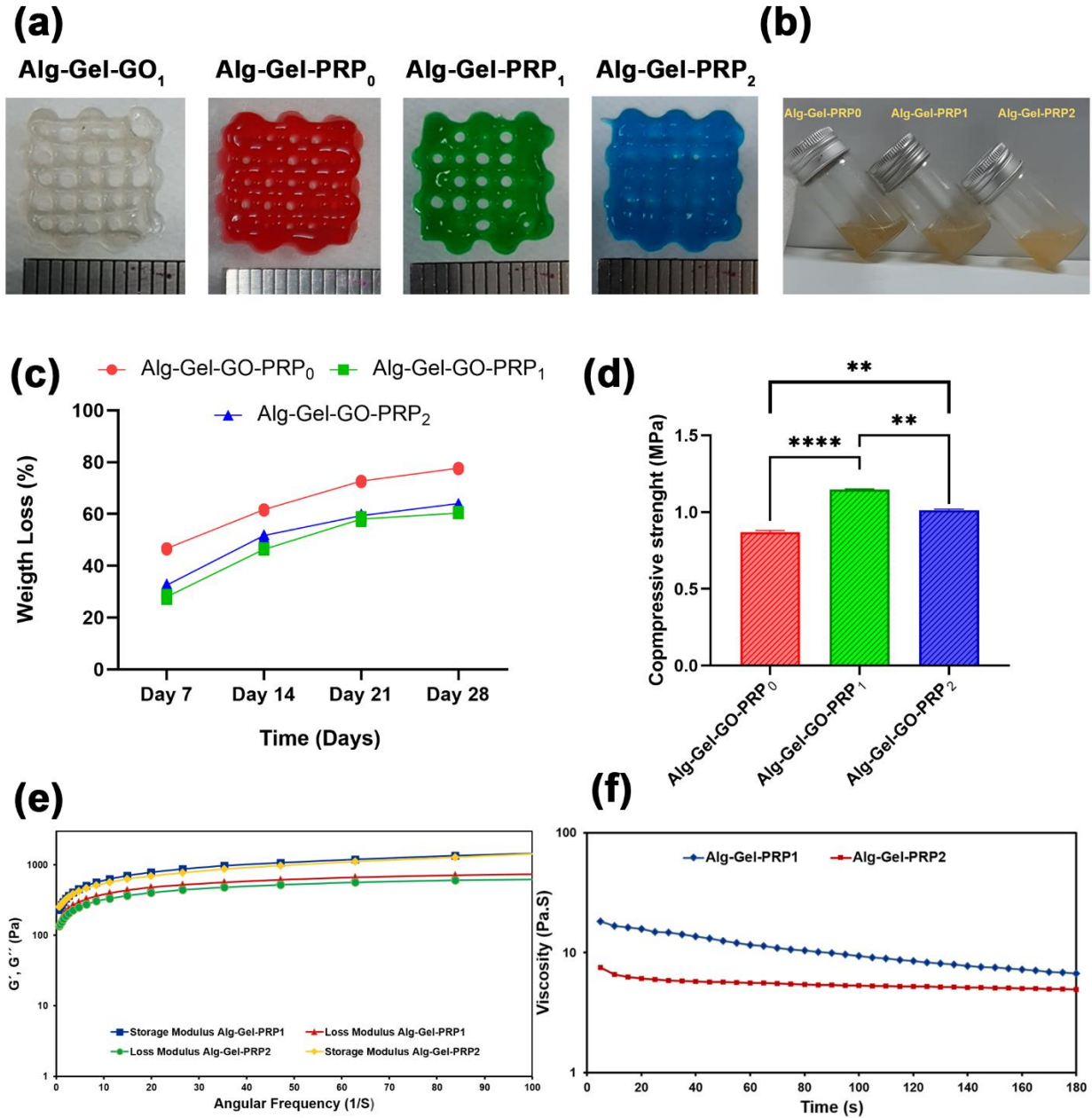

**Figure. 5. (a)** Printability assessment of bioink formulation. **(b)** The articular cartilage bioink composition with varying concentrations of PRP (0%, 1%, and 2%). **(c)** The degradation rate of 3D-printed multiphasic

scaffolds within 28 days. **(d)** Compressive strength of 3D-printed multiphasic scaffolds **(e)** Rheological evaluation of 3D-printed multiphasic scaffolds **(f)** Flow curves of viscosity of 3D-printed multiphasic scaffolds. Data expressed as mean ± SD (n = 3); ns: no significant difference, *p < 0.05, **p < 0.01, *** p < 0.001, **** p < 0.0001.

### 3.3. Biocompatibility characterization of multiphasic scaffold

### 3.3.1. Stem cell characterization

BM-MSCs were successfully extracted from rats' femurs and used for the biological evaluation of multiphasic scaffolds. **Figure. 6a** demonstrates a uniform population of cells that have a spindle-shaped morphology. The isolated cells were characterized by evaluating the expression of surface antigen markers, including CD73, CD105, CD90, CD45, and CD34 **(Figure. 6b$_{1-5}$)**. Flow cytometry analysis revealed that the BM-MSCs displayed high level expression of CD73 (98.6%), CD90 (96.9%), and CD105 (96.9%), but the levels of CD45 (0.487%) and CD34 (0.468%) expression were negative. The collected data align with earlier research and demonstrate that MSCs exhibit a specific array of surface antigen markers (such as CD73, CD90, and CD105) while lacking the surface antigen markers associated with hematopoietic cell lineages (such as CD45 and CD34) [23].

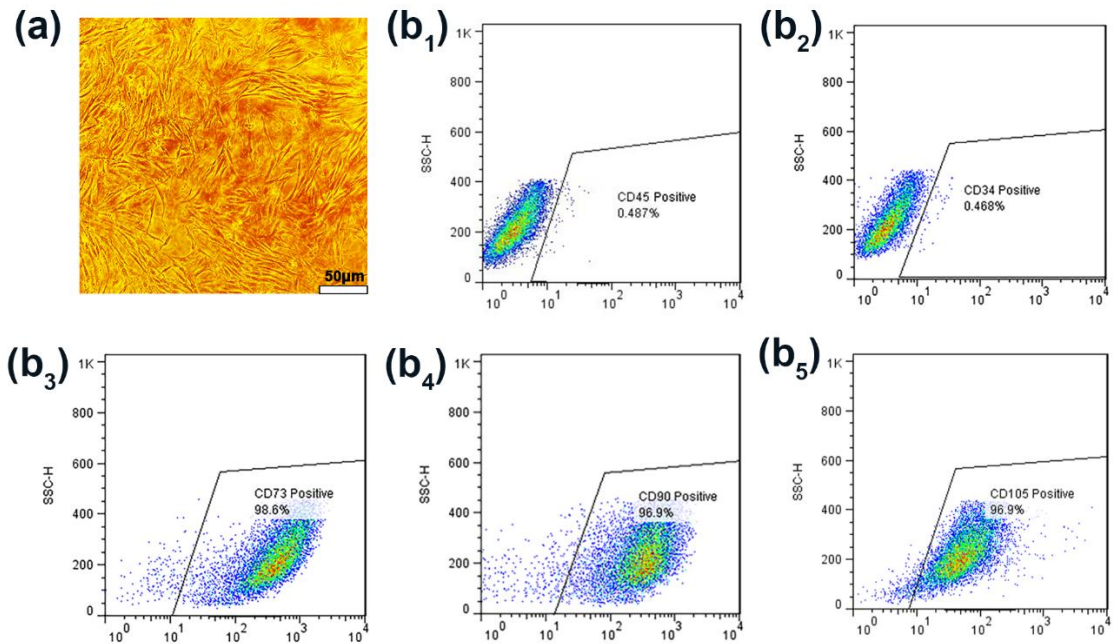
2929

**Figure. 6.** Results of the isolation and characterization of BM-MSCs. **(a)** Morphology of BM-MSCs in passage 3. Scatterplots for identification of BM-MSCs via flow cytometry. The expressions of **($b_1$)** CD45, **($b_2$)** CD34, **($b_3$)** CD73, **($b_4$)** CD90, and **($b_5$)** CD105 were detected in 0.487%, 0.468%, 98.6%, 96.9%, and 96.9% of the cells, respectively.

### 3.3.2. *In vitro* biocompatibility analysis of 3D-printed multiphasic scaffolds

To evaluate the viability of BM-MSCs seeded on 3D-printed multiphasic scaffolds with different PRP concentrations, MTT assays were performed. The results showed that on days 3 and 7, Alg-Gel-GO-PRP scaffolds with 1% PRP and 2% PRP significantly have higher cell proliferation as compared to the 0% PRP **(Figure. 7a)**. PRP offers a combination of natural growth factors and a hydrated extracellular matrix-like microenvironment that enhances cellular viability and proliferation. PRP promotes cellular proliferation and reduces cell death by inhibiting Bcl-2 expression and apoptosis [57,58]. These findings align with previous studies, which demonstrated enhanced cellular metabolic activity and proliferation in the presence of PRP [59]. For example, Singh *et al.* demonstrated that cells encapsulated within sodium alginate (SA), chitosan/chondroitin sulfate (PEC), silk fibroin (SF), and PRP scaffolds for cartilage tissue regeneration exhibited significantly higher metabolic activity and proliferation rates compared to PEC/SF and PEC/SF/SA scaffolds [57]. However, Choi *et al.* [60] reported a decrease in cell viability and proliferation at higher PRP concentrations. Similarly, Tavassoli-Hojjati *et al.* [61] found that while 5% PRP had the most significant effect on fibroblast proliferation after 3 days, increasing PRP concentration up to 50% led to reduce cell viability and proliferation. Therefore, Alg-Gel-GO-PRP$_2$ bioink was observed to be an optimal biomaterial for OC tissue regeneration.

**Figure. 7b** shows live/dead staining performed 72 hours after stem cell seeding on the scaffolds, highlighting a large number of viable cells (green) and a minimal presence of dead cells (red) across all groups. The different concentrations of PRP (0%, 1%, and 2% w/v) showed no obvious difference in cell viability, indicating great biocompatibility for all 3D-printed multiphasic scaffolds and no obvious cell mortality. In addition, the proliferation rate of viable cells in 2% PRP was higher than in the other groups (**Figure. 7b**). We did not see an increase in dead cells in the cell viability images at 72 hours and dead cells were likely degraded. All groups supported cell attachment and proper morphology; however, the multiphasic scaffolds with 2% PRP exhibited superior cell spreading and cell-cell interactions after 72 hours of culture. Increased PRP concentration enhanced cell-spreading capabilities, resulting in homogeneous dispersion of stem cells with spindle-shaped morphology. Stem cell attachment was assessed using FE-SEM after 48 hours (**Figure. 7c**). Results showed that the multiphasic scaffold with 2% PRP promoted favorable conditions for cell spreading and elongation, as PRP GFs like PDGF and TGF-β enhance cytoskeletal organization, improving cell-matrix interactions and overall morphology [62].

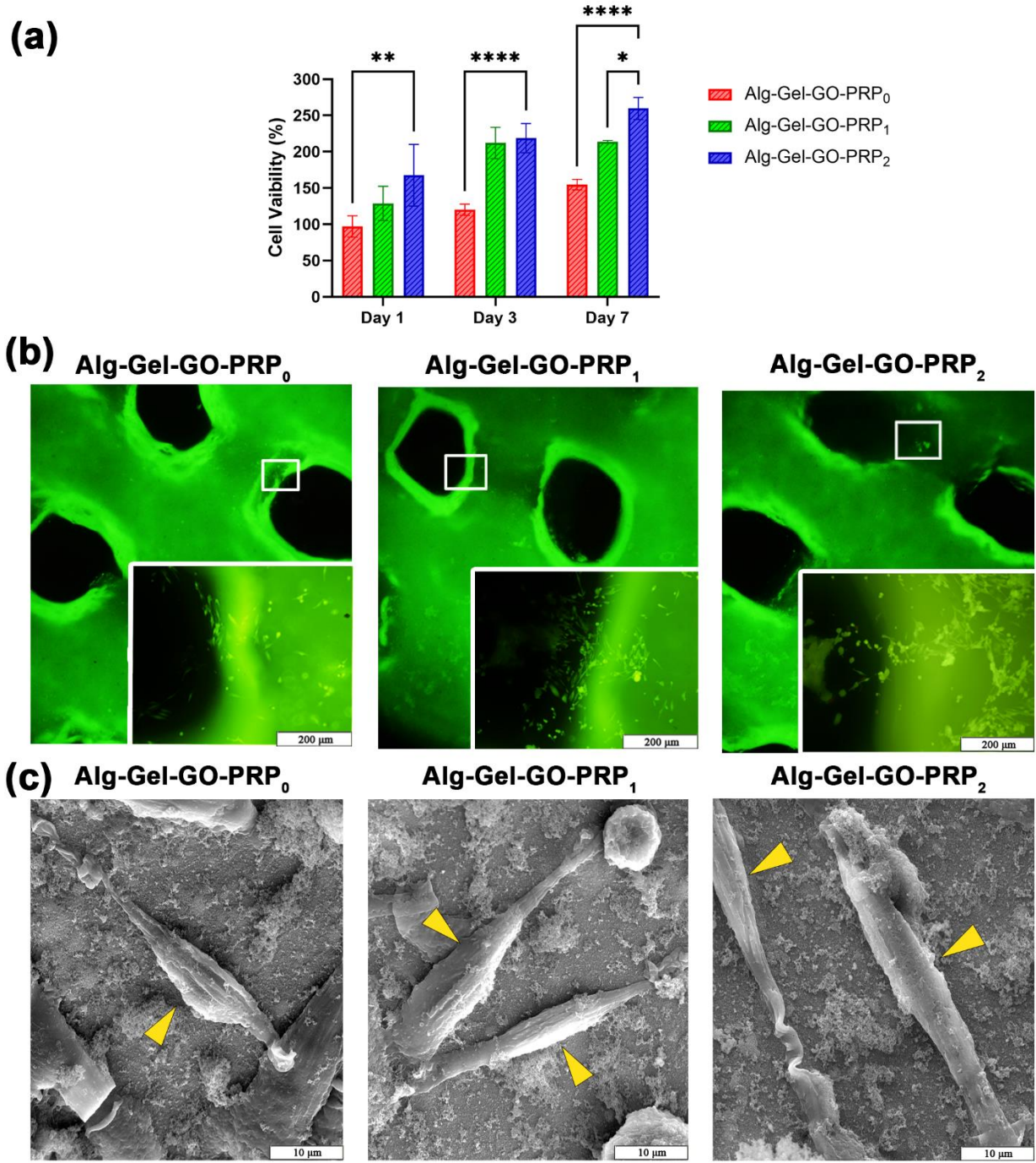

**Figure. 7.** Evaluation of stem cell viability. **(a)** MTT assay for assessing cell viability in 3D-printed multiphasic scaffolds; (ns, no significant difference; $^*p \leq 0.05$, *** $p \leq 0.001$, **** $p \leq 0.0001$). **(b)** Live-dead assay of BM-MSCs in different 3D-printed multiphasic scaffold groups (scale bar = 200 µm). **(c)** FE-SEM images of the adhesion of stem cells on the surface of different 3D-printed multiphasic scaffold

groups. The cells were well-attached to the different 3D-printed multiphasic scaffold groups (scale bar = 10 µm).

### 3.3.3. *In vitro* chondrogenic differentiation of BM-MSCs

The effect of PRP concentration on the chondrogenic differentiation of BM-MSCs on 3D-printed multiphasic scaffolds was evaluated by measuring the expression of cartilage-specific genes, including collagen type II and SOX9, along with bone-specific genes like collagen type I (**Figure. 8a$_{1-3}$**). The RT-PCR analysis revealed that the 3D-printed multiphasic scaffold groups containing PRP 2% strongly upregulated the chondrogenic markers SOX9 and collagen type II compared with the other groups (**P ≤ 0.01 with 1% PRP, ***P ≤ 0.001 without PRP). Previous research indicated that a high concentration of 10% PRP in the differentiation medium could induce adipose-derived MSCs to differentiate into chondrocytes [63]. In addition, lyophilized PRP powder plays remarkable roles as a storage vehicle of GFs, including TGF-β and PDGF, as demonstrated by ELISA tests. TGF-β and PDGF are well known to induce the MSC chondroblast differentiation and the accumulation of cartilage ECM [64]. Conversely, the mRNA level of collagen type I (a positive osteogenesis marker) was downregulated in the multiphasic scaffold group with 2% PRP compared to other chondrogenic markers like collagen type II and Sox9. The previous study has demonstrated that adding 5% PRP to the culture medium of adipose-derived MSCs enhances osteogenic marker expression [65]. However, our research indicates that incorporating 2% PRP in a 3D-printed multiphasic scaffold group increases chondrogenic marker expression more than osteogenic marker expression (***P ≤ 0.001) (**Figure. 8a$_4$**). Moreover, *in vitro* chondrogenesis of BM-MSCs in the 3D-printed multiphasic scaffold was further evaluated by histological assessment of ECM deposition using Alcian Blue staining (**Figure. 8b**). With increasing GFs concentration, PRP-loaded scaffold groups showed greater positive staining for Alcian Blue in the matrix surrounding the stem cells. Alcian Blue staining revealed that 2% PRP in 3D-printed scaffolds increased staining intensity, indicating higher GAGs content in the ECM around the stem cells compared with the other groups [57] (**Figure. 8b**). Notably, BM-MSCs on scaffolds with 2% PRP exhibited significantly enhanced chondrogenic gene expression and generated a cartilage-like ECM. Our findings showed that 2% PRP improved cell responsiveness and effectively promoted chondrogenic differentiation, along with cartilage ECM regeneration of BM-MSCs in a biomimetic environment.

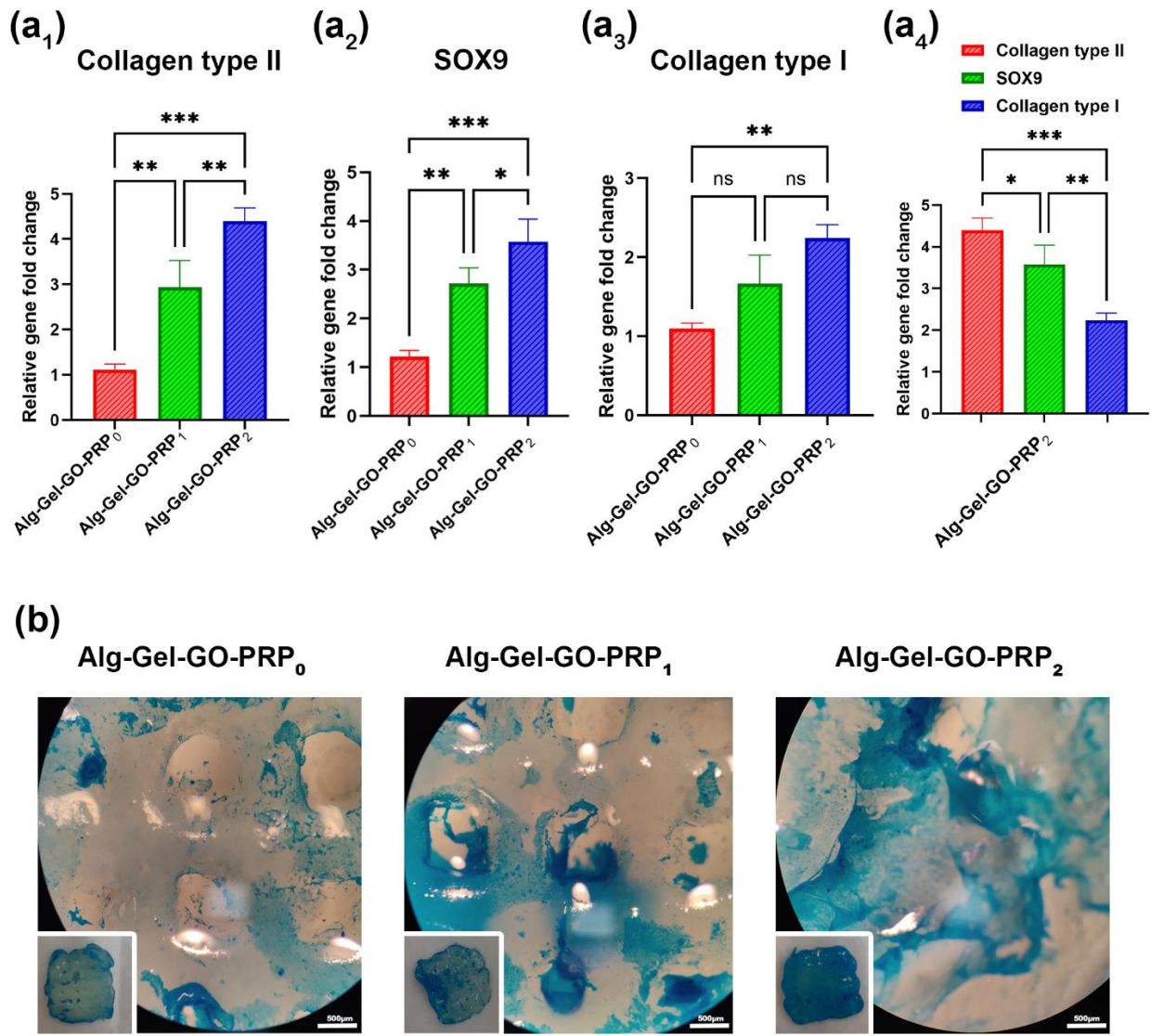

**Figure. 8.** (**a₁₋₄**) Gene expression levels of chondrogenic markers (collagen type II, SOX9) and the osteogenic marker (collagen type I) in different multiphasic scaffold groups by BM-MSCs. The values are presented as fold changes and normalized against the GAPDH reference value (n = 3); ns, no significant difference; $^*p \leq 0.05$, *** $p \leq 0.001$, **** $p \leq 0.0001$. (**b**) Alcian Blue staining of BM-MSCs following 3 weeks of chondrogenic differentiation on the surface of the multiphasic scaffolds (scale bar = 500 µm).

## 4. Conclusion

This study utilized extrusion-based multi-nozzle 3D printing to create tri-layered multiphasic scaffolds that mimicked OC tissue. First, we added varying concentrations of GO to the Alg-Gel bioink as a subchondral bone layer and investigated its chemical composition, swelling, rheological and mechanical properties, and biocompatibility. The optimal concentration for printing the subchondral layer construct and preserving its density in the other multiphasic scaffold layer was Alg-Gel-GO$_1$. Then, Alg-Gel-GO bioink was used to print multiphasic scaffolds with varying PRP concentrations. The degradation, rheological, mechanical, and printability properties of the bioinks were further investigated, and scaffold biocompatibility was tested *in vitro* using BM-MSCs. We found that all bioinks increased BM-MSCs chondrogenic development after 3 weeks in chondrogenic media. The bioink composed of Alg-Gel-GO-PRP$_2$ was the best for chondrogenic gene expression and GAGs deposition. According to our experiment, the 3D-printed Alg-Gel-GO-PRP$_2$ is a good option for OCTE due to its biocompatibility. Our findings show that 3D-printed OC scaffolds with multiphasic structures promote osteogenesis and chondrogenesis and have considerable potential for OC repair and regeneration.

**Acknowledgment**
Iran University of Medical Sciences, Cellular and Molecular Research Center, with grant No. 24717.

**Author Contributions**
**Faezeh Ghobadi:** Conceptualization, Methodology, Formal Analysis, Visualization, and Writing Original Draft. **Maryam Mohammadi:** Resources. **Rooja Kalantarzadeh:** Assistance with experimental work. **Ehsan Lotfi:** Formal Analysis. **Shokoufeh Borhan:** Resources. **Narendra Pal Singh Chauhan:** Reviewing and Editing. **Ghazaleh Salehi:** Supervision, Methodology, Writing Original Draft, and Resources. **Sara Simorgh:** Supervision, Project Administration, Reviewing and Editing, and Resources.

**Data Availability**
The datasets collected and/or examined in this research can be obtained from the author responsible for the study upon a reasonable request.

**Competing Interests**
The authors declare no competing interests.

**References**

[1] S. Ansari, S. Khorshidi, A. Karkhaneh, *Acta Biomater* **2019**, *87*, 41.

[2] J. Liu, L. Li, H. Suo, M. Yan, J. Yin, J. Fu, *Mater Des* **2019**, *171*, 107708.

[3] R. Longley, A. M. Ferreira, P. Gentile, *International Journal of Molecular Sciences 2018, Vol. 19, Page 1755* **2018**, *19*, 1755.

[4] D. Algul, H. Sipahi, A. Aydin, F. Kelleci, S. Ozdatli, F. G. Yener, *Int J Biol Macromol* **2015**, *79*, 363.

[5] B. Zhang, Y. Luo, L. Ma, L. Gao, Y. Li, Q. Xue, H. Yang, Z. Cui, *Biodes Manuf* **2018**, *1*, 2.

[6] S. Jia, J. Wang, T. Zhang, W. Pan, Z. Li, X. He, C. Yang, Q. Wu, W. Sun, Z. Xiong, D. Hao, *ACS Appl Mater Interfaces* **2018**, *10*, 20296.

[7] A. M. Yousefi, M. E. Hoque, R. G. S. V. Prasad, N. Uth, *J Biomed Mater Res A* **2015**, *103*, 2460.

[8] J. Jang, J. Y. Park, G. Gao, D. W. Cho, *Biomaterials* **2018**, *156*, 88.

[9] C. Tai, S. Bouissil, E. Gantumur, M. S. Carranza, A. Yoshii, S. Sakai, G. Pierre, P. Michaud, C. Delattre, *Applied Sciences 2019, Vol. 9, Page 2596* **2019**, *9*, 2596.

[10] J. Jia, D. J. Richards, S. Pollard, Y. Tan, J. Rodriguez, R. P. Visconti, T. C. Trusk, M. J. Yost, H. Yao, R. R. Markwald, Y. Mei, *Acta Biomater* **2014**, *10*, 4323.

[11] R. Yao, R. Zhang, J. Luan, F. Lin, *Biofabrication* **2012**, *4*, DOI 10.1088/1758-5082/4/2/025007.

[12] F. Olate-Moya, L. Arens, M. Wilhelm, M. A. Mateos-Timoneda, E. Engel, H. Palza, *ACS Appl Mater Interfaces* **2020**, *12*, 4343.

[13] S. D. Purohit, R. Bhaskar, H. Singh, I. Yadav, M. K. Gupta, N. C. Mishra, *Int J Biol Macromol* **2019**, *133*, 592.

[14] M. Yazdimamaghani, D. Vashaee, S. Assefa, M. Shabrangharehdasht, A. T. Rad, M. A. Eastman, K. J. Walker, S. V. Madihally, G. A. Köhler, L. Tayebi, *Mater Sci Eng C Mater Biol Appl* **2014**, *39*, 235.

[15] R. S. Tuan, A. F. Chen, B. A. Klatt, *J Am Acad Orthop Surg* **2013**, *21*, 303.

[16] K. Okuda, T. Kawase, M. Momose, M. Murata, Y. Saito, H. Suzuki, L. F. Wolff, H. Yoshie, *J Periodontol* **2003**, *74*, 849.

[17] B. Johnstone, T. M. Hering, A. I. Caplan, V. M. Goldberg, J. U. Yoo, *Exp Cell Res* **1998**, *238*, 265.

[18] M. Y. Mollajavadi, M. Saadatmand, F. Ghobadi, *Iranian Polymer Journal (English Edition)* **2023**, *32*, 599.

[19] F. Heidari, M. Saadatmand, S. Simorgh, *Int J Biol Macromol* **2023**, *253*, DOI 10.1016/J.IJBIOMAC.2023.127041.

[20] A. A. Menarbazari, A. Mansoori-Kermani, S. Mashayekhan, A. Soleiymani, *Int J Biol Macromol* **2024**, *265*, DOI 10.1016/J.IJBIOMAC.2024.130827.

[21] M. Ansarizadeh, S. Mashayekhan, M. Saadatmand, *Int J Biol Macromol* **2019**, *125*, 383.

[22] L. Ouyang, R. Yao, Y. Zhao, W. Sun, *Biofabrication* **2016**, *8*, DOI 10.1088/1758-5090/8/3/035020.

[23] N. Alasvand, S. Simorgh, M. Malekzadeh Kebria, A. Bozorgi, S. Moradi, V. Hosseinpour Sarmadi, K. Ebrahimzadeh, N. Amini, F. Kermani, S. Kargozar, P. Brouki Milan, *Open Ceramics* **2023**, *14*, 100358.

[24] F. Ghobadi, M. Saadatmand, S. Simorgh, P. Brouki Milan, *Scientific Reports 2024 15:1* **2025**, *15*, 1.

[25] D. Kilian, T. Ahlfeld, A. R. Akkineni, A. Bernhardt, M. Gelinsky, A. Lode, *Scientific Reports 2020 10:1* **2020**, *10*, 1.

[26] E. Lotfi, A. Kholghi, F. Golab, A. Mohammadi, M. Barati, *Pathol Res Pract* **2024**, *255*, DOI 10.1016/J.PRP.2024.155187.

[27] X. Zhou, M. Nowicki, H. Cui, W. Zhu, X. Fang, S. Miao, S. J. Lee, M. Keidar, L. G. Zhang, *Carbon N Y* **2017**, *116*, 615.

[28] L. T. Somasekharan, N. Kasoju, R. Raju, A. Bhatt, *Bioengineering (Basel)* **2020**, *7*, 1.

[29] J. Zhang, H. Eyisoylu, X. H. Qin, M. Rubert, R. Müller, *Acta Biomater* **2021**, *121*, 637.

[30] X. Hu, Y. Man, W. Li, L. Li, J. Xu, R. Parungao, Y. Wang, S. Zheng, Y. Nie, T. Liu, K. Song, *Polymers (Basel)* **2019**, *11*, DOI 10.3390/POLYM11101601.

[31] G. Choe, S. Oh, J. M. Seok, S. A. Park, J. Y. Lee, *Nanoscale* **2019**, *11*, 23275.

[32] P. Gupta, A. Agrawal, K. Murali, R. Varshney, S. Beniwal, S. Manhas, P. Roy, D. Lahiri, *Materials Science and Engineering: C* **2019**, *97*, 539.

[33] S. Sinha, A. Astani, T. Ghosh, P. Schnitzler, B. Ray, *Phytochemistry* **2010**, *71*, 235.

[34] J. Kim, J. Park, G. Choe, S. I. Jeong, H. S. Kim, J. Y. Lee, *Adv Healthc Mater* **2024**, *13*, 2400142.

[35] C. Jiao, T. Li, J. Wang, H. Wang, X. Zhang, X. Han, Z. Du, Y. Shang, Y. Chen, *J Polym Environ* **2020**, *28*, 1492.

[36] M. C. F. Costa, V. S. Marangoni, P. R. Ng, H. T. L. Nguyen, A. Carvalho, A. H. Castro Neto, *Nanomaterials 2021, Vol. 11, Page 551* **2021**, *11*, 551.

[37] N. Khoshnood, A. Zamanian, *J Appl Polym Sci* **2022**, *139*, DOI 10.1002/APP.52227.

[38] A. Kumar, K. C. Nune, R. D. K. Misra, *J Tissue Eng Regen Med* **2018**, *12*, 1133.

[39] C. Sharma, A. K. Dinda, P. D. Potdar, C. F. Chou, N. C. Mishra, *Materials Science and Engineering: C* **2016**, *64*, 416.

[40] Kenry, W. C. Lee, K. P. Loh, C. T. Lim, *Biomaterials* **2018**, *155*, 236.

[41] D. Ramani, T. P. Sastry, *Cellulose* **2014**, *21*, 3585.

[42] S. D. Purohit, H. Singh, R. Bhaskar, I. Yadav, S. Bhushan, M. K. Gupta, A. Kumar, N. C. Mishra, *Front Mater* **2020**, *7*, 511489.

[43] Y. Zhu, S. Murali, W. Cai, X. Li, J. W. Suk, J. R. Potts, R. S. Ruoff, *Adv Mater* **2010**, *22*, 3906.

[44] E. Jain, N. Chinzei, A. Blanco, N. Case, L. J. Sandell, S. Sell, M. F. Rai, S. P. Zustiak, *J Orthop Res* **2019**, *37*, 2401.

[45] J. Zhang, Q. Luo, Q. Hu, T. Zhang, J. Shi, L. Kong, D. Fu, C. Yang, Z. Zhang, *Acta Pharm Sin B* **2023**, *13*, 4318.

[46] D. Chen, P. Chang, P. Ding, S. Liu, Q. Rao, O. V. Okoro, L. Wang, L. Fan, A. Shavandi, L. Nie, *Heliyon* **2023**, *9*, e14349.

[47] A. Castillo-Macías, J. Zavala, W. Ortega-Lara, J. E. Valdez-García, S. M. García-Herrera, *Clin Ophthalmol* **2023**, *17*, 3787.

[48] S. F. Braga, E. Trovatti, R. A. de Carvalho, A. J. F. de Carvalho, M. R. da C. Iemma, A. C. Amaral, *Brazilian Archives of Biology and Technology* **2020**, *63*, e20190003.

[49] Y. P. Singh, J. C. Moses, A. Bandyopadhyay, B. B. Mandal, *Adv Healthc Mater* **2022**, *11*, 2200209.

[50] T. H. Lin, H. C. Wang, W. H. Cheng, H. C. Hsu, M. L. Yeh, *Int J Mol Sci* **2019**, *20*, 326.

[51] M. Zhao, J. Wang, J. Zhang, J. Huang, L. Luo, Y. Yang, K. Shen, T. Jiao, Y. Jia, W. Lian, J. Li, Y. Wang, Q. Lian, D. Hu, *Mater Today Bio* **2022**, *16*, 100334.

[52] B. Sarker, D. G. Papageorgiou, R. Silva, T. Zehnder, F. Gul-E-Noor, M. Bertmer, J. Kaschta, K. Chrissafis, R. Detsch, A. R. Boccaccini, *J Mater Chem B* **2014**, *2*, 1470.

[53] Z. Li, X. Zhang, T. Yuan, Y. Zhang, C. Luo, J. Zhang, Y. Liu, W. Fan, *https://home.liebertpub.com/tea* **2020**, *26*, 886.

[54] D. Wu, A. Spanou, A. Diez-Escudero, C. Persson, *J Mech Behav Biomed Mater* **2020**, *103*, 103608.

[55] S. Tang, L. Wang, Y. Zhang, F. Zhang, *Front Bioeng Biotechnol* **2022**, *10*, DOI 10.3389/FBIOE.2022.887454.

[56] N. KhaliliJafarabad, A. Behnamghader, M. T. Khorasani, M. Mozafari, *Biotechnol Appl Biochem* **2022**, *69*, 534.

[57] B. N. Singh, A. Nallakumarasamy, S. Sinha, A. Rastogi, S. P. Mallick, S. Divakar, P. Srivastava, *Int J Biol Macromol* **2022**, *203*, 389.

[58] Y. Fukaya, M. Kuroda, Y. Aoyagi, S. Asada, Y. Kubota, Y. Okamoto, T. Nakayama, Y. Saito, K. Satoh, H. Bujo, *Exp Mol Med* **2012**, *44*, 330.

[59] X. Gao, L. Gao, T. Groth, T. Liu, D. He, M. Wang, F. Gong, J. Chu, M. Zhao, *J Biomed Mater Res A* **2019**, *107*, 2076.

[60] B. H. Choi, S. J. Zhu, B. Y. Kim, J. Y. Huh, S. H. Lee, J. H. Jung, *Int J Oral Maxillofac Surg* **2005**, *34*, 420.

[61] S. Tavassoli-Hojjati, M. Sattari, T. Ghasemi, R. Ahmadi, A. Mashayekhi, *Eur J Dent* **2016**, *10*, 469.

[62] M. Fukui, F. Lai, M. Hihara, T. Mitsui, Y. Matsuoka, Z. Sun, S. Kunieda, S. Taketani, T. Odaka, K. Okuma, N. Kakudo, *Hum Cell* **2024**, *37*, 181.

[63] A. Barlian, H. Judawisastra, N. M. Alfarafisa, U. A. Wibowo, I. Rosadi, *PeerJ* **2018**, *6*, DOI 10.7717/PEERJ.5809.

[64] S. Elder, J. Thomason, *Open Orthop J* **2014**, *8*, 78.

[65] M. Kazem-Arki, M. Kabiri, I. Rad, N. H. Roodbari, H. Hosseinpoor, S. Mirzaei, K. Parivar, H. Hanaee-Ahvaz, *Cytotechnology* **2018**, *70*, 1487.